\documentclass[12pt]{spieman}  % 12pt font required by SPIE;
\usepackage{amsmath,amsfonts,amssymb}
\usepackage{graphicx}
\usepackage{setspace}
\usepackage{tocloft}
\usepackage{lineno}
\usepackage[utf8]{inputenc}
\usepackage[english]{babel}
\usepackage{csquotes}
\usepackage{microtype}
\usepackage{isotope}
\usepackage{subcaption}
\usepackage{orcidlink}
\usepackage{xcolor}
\usepackage{booktabs}
\usepackage[capitalize,noabbrev]{cleveref}
\DeclareCaptionOption{class=article}
\usepackage{caption}

\usepackage{nowidow}
\usepackage{xspace} % simplify spacing for abbreviations

\def\eg{{\it e.g.}\xspace}

\def\spider{{\sc Spider}\xspace}
\def\Spider{{\sc Spider}\xspace}

\def\planck{{\it Planck}\xspace}

\def\krj{K$_{\text{RJ}}$\xspace}

\def\mukcmbrts{$\mu$K$_{\text{CMB}}\sqrt{s}$\xspace}

\def\sims{$\sim$\xspace}
\title{In-Flight Performance of SPIDER's 280 GHz Receivers}

\author[a,b,c]{E.~C.~Shaw}

\author[d]{P.~A.~R.~Ade}

\author[e]{S.~Akers}
\author[f]{M.~Amiri}
\author[g]{J.~Austermann}
\author[g]{J.~Beall}
\author[g]{D.~T.~Becker}
\author[h]{S.~J.~Benton}
\author[h]{A.~S.~Bergman}
\author[i,j]{J.~J.~Bock}
\author[k]{J.~R.~Bond}
\author[l]{S.~A.~Bryan}
\author[m,n]{H.~C.~Chiang}
\author[o]{C.~R.~Contaldi}
\author[p]{R.~S.~Domagalski}
\author[i,j]{O.~Dor\'e}
\author[g]{S.~M.~Duff}
\author[q,h]{A.~J.~Duivenvoorden}
\author[r]{H.~K.~Eriksen}
\author[s]{M.~Farhang}
\author[c]{J.~P.~Filippini}
\author[t]{L.~M.~Fissel}
\author[h]{A.~A.~Fraisse}
\author[a,b,u]{K.~Freese}
\author[r]{M.~Galloway}
\author[v]{A.~E.~Gambrel}
\author[w]{N.~N.~Gandilo}
\author[x]{K.~Ganga}
\author[c]{S.~M.~Gibbs}
\author[h]{S.~Gourapura}
\author[g]{A.~Grigorian}
\author[c,y]{R.~Gualtieri}
\author[z,u]{J.~E.~Gudmundsson}
\author[f]{M.~Halpern}
\author[p]{J.~Hartley}
\author[1]{M.~Hasselfield}
\author[g]{G.~Hilton}
\author[j]{W.~Holmes}
\author[i]{V.~V.~Hristov}
\author[k]{Z.~Huang}
\author[g]{J.~Hubmayr}
\author[2,3]{K.~D.~Irwin}
\author[h]{W.~C.~Jones}
\author[c]{A.~Kahn}
\author[h]{Z.~D.~Kermish}
\author[e]{C.~King}
\author[2]{C.~L.~Kuo}
\author[c]{A.~R.~Lennox}
\author[p,4]{J.~S.-Y.~Leung}
\author[h,5]{S.~Li}
\author[h]{T.~V.~Luu}
\author[i]{P.~V.~Mason}
\author[e]{J.~May}
\author[j]{K.~Megerian}
\author[i]{L.~Moncelsi}
\author[i]{T.~A.~Morford}
\author[e]{J.~M.~Nagy}
\author[c]{R.~Nie}
\author[p,6]{C.~B.~Netterfield}
\author[k]{M.~Nolta}
\author[c]{B.~Osherson}
\author[e,p,7]{I.~L.~Padilla}
\author[8,v]{A.~S.~Rahlin}
\author[i,j]{S.~Redmond}
\author[g]{C.~Reintsema}
\author[9]{L.~J.~Romualdez}
\author[e]{J.~E.~Ruhl}
\author[j]{M.~C.~Runyan}
\author[k]{J.~A.~Shariff}
\author[h]{C.~Shiu}
\author[10]{J.~D.~Soler}
\author[h]{X.~Song}
\author[h]{S.~Tartakovsky}
\author[r]{H.~Thommesen}
\author[i,j]{A.~Trangsrud}
\author[d]{C.~Tucker}
\author[j]{A.~D.~Turner}
\author[g]{J.~Ullom}
\author[h]{J.~F.~van~der~List}
\author[g]{J.~Van~Lanen}
\author[g]{M.~R.~Vissers}
\author[j]{A.~C.~Weber}
\author[r]{I.~K.~Wehus}
\author[e]{S.~Wen}
\author[f]{D.~V.~Wiebe}
\author[2,3]{E.~Y.~Young}
\affil[a]{Department of Physics, The University of Texas at Austin, Austin, TX 78712, USA}
\affil[b]{Weinberg Institute for Theoretical Physics, Texas Center for Cosmology and Astroparticle Physics, Austin, TX 78712, USA}
\affil[c]{Department of Physics, University of Illinois Urbana-Champaign, 1110 W Green St, Urbana, IL 61801, USA}
\affil[d]{School of Physics and Astronomy, Cardiff University, Cardiff CF24 3AA, UK}
\affil[e]{Department of Physics, Case Western Reserve University, 10900 Euclid Ave, Cleveland, OH 44106, USA}
\affil[f]{Department of Physics and Astronomy, University of British Columbia, Vancouver, BC V6T 1Z1, Canada}
\affil[g]{National Institute of Standards and Technology, 325 Broadway Mailcode 817.03, Boulder, CO 80305, USA}
\affil[h]{Department of Physics, Princeton University, Jadwin Hall, Princeton, NJ 08544, USA}
\affil[i]{Division of Physics, Mathematics and Astronomy, California Institute of Technology, Pasadena, CA 91125, USA}
\affil[j]{Jet Propulsion Laboratory, Pasadena, CA 91011, USA}
\affil[k]{Canadian Institute for Theoretical Astrophysics, University of Toronto, Toronto, ON M5S 3H4, Canada}
\affil[l]{School of Earth and Space Exploration, Arizona State University, Tempe, AZ 85283, USA}
\affil[m]{Department of Physics, McGill University, Montreal, QC H3A 2T8, Canada}
\affil[n]{School of Mathematics, Statistics and Computer Science, University of KwaZulu-Natal, Durban, South Africa}
\affil[o]{Blackett Laboratory, Imperial College London, London SW7 2AZ, UK}
\affil[p]{Department of Astronomy and Astrophysics, University of Toronto, Toronto, ON M5S 3H4, Canada}
\affil[q]{Center for Computational Astrophysics, Flatiron Institute, New York, NY 10010, USA}
\affil[r]{Institute of Theoretical Astrophysics, University of Oslo, NO-0315 Oslo, Norway}
\affil[s]{Department of Physics, Shahid Beheshti University, 1983969411, Tehran Iran}
\affil[t]{Department of Physics, Engineering Physics and Astronomy, Queen's University, Kingston, ON K7L 3N6, Canada}
\affil[u]{The Oskar Klein Centre for Cosmoparticle Physics, Department of Physics, Stockholm University, AlbaNova, SE-106 91 Stockholm, Sweden}
\affil[v]{Kavli Institute for Cosmological Physics, University of Chicago, Chicago, IL 60637, USA}
\affil[w]{Steward Observatory, Tuscon, AZ 85721, USA}
\affil[x]{Universit\'e de Paris, CNRS, AstroParticule et Cosmologie, F-75013 Paris, France}
\affil[y]{Department of Physics and Astronomy, Northwestern University, Evanston, IL 60208, USA}
\affil[z]{Science Institute, University of Iceland, 107 Reykjavik, Iceland}
\affil[1]{Department of Astronomy and Astrophysics, Pennsylvania State University, University Park, PA 16802, USA}
\affil[2]{Department of Physics, Stanford University, Stanford, CA 94305, USA}
\affil[3]{SLAC National Accelerator Laboratory, Menlo Park, CA 94025, USA}
\affil[4]{Dunlap Institute for Astronomy and Astrophysics, University of Toronto, Toronto, ON M5S 3H4, Canada}
\affil[5]{Department of Mechanical and Aerospace Engineering, Princeton University, Engineering Quadrangle, Princeton, NJ 08544, USA}
\affil[6]{Department of Physics, University of Toronto, Toronto, ON M5S 3H4, Canada}
\affil[7]{Department of Physics and Astronomy, Johns Hopkins University, Baltimore, MD 21218, USA}
\affil[8]{Department of Astronomy and Astrophysics, University of Chicago, Chicago, IL 60637, USA}
\affil[9]{University of Toronto Institute for Aerospace Studies, Toronto, ON M5S 3H4, Canada}
\affil[10]{Max-Planck-Institute for Astronomy, Konigstuhl 17, 69117, Heidelberg, Germany}

\cftpagenumbersoff{figure}
\cftpagenumbersoff{table} 
\begin{document} 
\maketitle

\begin{abstract}
\singlespacing
\spider is a balloon-borne instrument designed to map the cosmic microwave background at degree-angular scales in the presence of Galactic foregrounds.
\spider has mapped a large sky area in the Southern Hemisphere using more than 2000 transition-edge sensors (TESs) during two NASA Long Duration Balloon flights above the Antarctic continent.
During its first flight in January 2015, \spider observed in the 95~GHz and 150~GHz frequency bands, setting constraints on the B-mode signature of primordial gravitational waves.
Its second flight in the 2022-23 season added new receivers at 280~GHz, each using an array of TESs coupled to the sky through feedhorns formed from stacks of silicon wafers.
These receivers are optimized to produce deep maps of polarized Galactic dust emission over a large sky area, providing a unique data set with lasting value to the field.
We describe the instrument's performance during \Spider's second flight, focusing on the performance of the 280~GHz receivers.
We include details on the flight, in-band optical loading at float, and an early analysis of detector noise. 
\end{abstract}

% Include a list of up to six keywords after the abstract
%\keywords{\spider, cosmic microwave background, feedhorn coupled transition-edge sensors, scientific ballooning, 280~GHz cosmology}

% Include email contact information for corresponding author
{\noindent \footnotesize\textbf{*}Elle C. Shaw,  \linkable{elle.shaw@utexas.edu} }

\begin{spacing}{1}  

\section{Introduction}
\label{sec:intro}  % \label{} allows reference to this section
\footnote{
This is the author’s version of the manuscript [\citenum{InflightSpider2JATIS}] accepted for publication in \textit{Journal of Astronomical Telescopes, Instruments, and Systems (JATIS)}. The final version is Copyright 2024 SPIE, and is available at https://doi.org/10.1117/1.JATIS.10.4.044012}
\Spider is a balloon-borne instrument specifically designed to advance the search for the inflationary signature of primordial gravitational waves in the \textit{B}-mode polarization of the cosmic microwave background (CMB) in the presence of Galactic foregrounds~\cite{Filippini_2010,  Ade_2015, Fraisse_2013, Gualtieri_2018}.
The inflationary \textit{B}-mode polarization signal is expected to be extremely faint, with anisotropies  $\ll 1~\mu \text{K}$. 
Experimental detection of such a signal requires extremely sensitive detectors, tight control over polarized instrumental systematics, and effective strategies to mitigate the effects from atmospheric and Galactic sources.
Earth's millimeter-wave absorptive and emissive atmosphere attenuates the CMB signal, increases optical loading, and produces fake signals from fluctuations in atmospheric brightness. 
The two primary sources of Galactic foreground polarization, synchrotron radiation and thermal emission from dust, overshadow the polarized CMB signals at all frequencies when the brightness is averaged across the full sky~\cite{collaboration2018planck}. 
Even away from the Galactic plane in relatively `clean' regions of the sky, the diffuse thermal dust emission is significant and requires proper characterization to distinguish from any CMB signal. 
The isolation of CMB signals from foregrounds requires high-quality observations in multiple frequency bands, where the different spectral behaviors of each component can, in principle, be leveraged to separate them.

The amplitude of primordial gravitational waves is characterized by the tensor-to-scalar ratio $r$.
\spider's mission is to set an upper limit on $r$ at the few-percent level while producing high-quality maps of polarized Galactic dust emission.
\Spider has mapped the polarization of the millimeter sky in three frequency bands centered near 95~GHz, 150~GHz, and 280~GHz during two NASA Long Duration Balloon (LDB) flights above the Antarctic continent.
At an altitude of $\sim$35~km, the ballooning platform allows for observations over a wide sky area with minimal contamination from atmospheric emission.
Under such conditions, we can maximize detector sensitivity to make high-fidelity maps in a matter of days.

\begin{figure}[tbp]
\centering
\includegraphics[width=.75\textwidth]{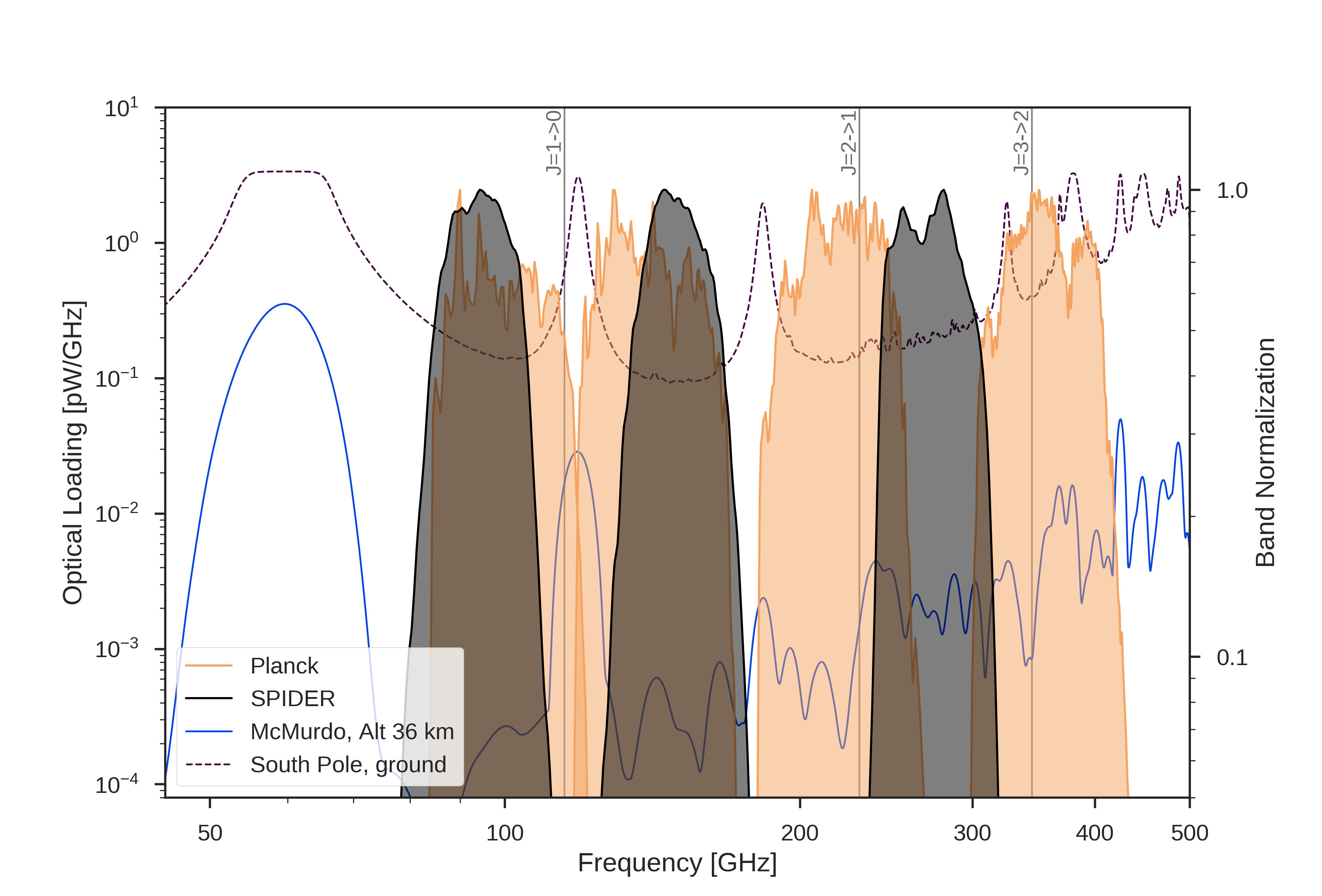}% B
\caption{Measured spectral bandpasses from \spider-2 (black) and \planck HFI (orange) superimposed over the atmospheric optical loading for an idealized single-moded detector located at the South Pole (dash) and at 36 km above McMurdo Station, Antarctica (blue), as calculated using the ``am'' atmospheric modeling software suite~\cite{am_software,Bergman_thesis}. Rotational transition spectral lines for Galactic CO are also shown. \spider's 280~GHz band provides an independent measurement of polarized Galactic dust emission at frequencies that are difficult to access from the ground. The bandpasses shown for \spider in the 90 and 150 GHz bands are the average bandpass for the detectors flown on \spider-2, and the 280 GHz bandpass is the average of the two of the three receivers (See \cref{subsec:noisecalib} for more details).}
\label{fig:ObservationBands}
\end{figure}

The detailed shape of \spider's measured bands is shown in~\cref{fig:ObservationBands}, together with the polarized bands of the \planck High Frequency Instrument (HFI).
Atmospheric loading brightness for observation locations at the South Pole and at an altitude of 36~km is superimposed on the frequency bands\cite{am_software,Bergman_thesis}, showing the placement of \spider's observation bands between absorption lines.
\spider's 280~GHz band complements the 95 and 150~GHz data sets by providing high-contrast dust maps with a shorter lever arm than the \planck 353~GHz band to extrapolate dust models to lower frequencies.

In January 2015, \spider completed its first flight from the LDB Facility at McMurdo Station, Antarctica.
For clarity, we refer to this flight as `\spider-1'.
During a 16-day flight, \spider-1 mapped approximately 10\% of the sky with three receivers tuned to 95~GHz and three to 150~GHz.
The large sky area allows us to sample the angular power spectrum over a wide range of multipoles, $35\lesssim \ell\lesssim250$, covering the relevant multipole range for the expected peak in the primordial \textit{B}-mode signal near $\ell\sim80$.
\spider-2 is the second flight of the \spider experiment, which targeted the same sky region during another 16-day flight in December 2022 -- January 2023.
The second flight redeployed three \spider-1 receivers (two at 95~GHz and one at 150~GHz) and incorporated three novel 280~GHz receivers.
The \spider 280~GHz receivers are vastly more sensitive than those from any prior experiment in this band.
The maps from \spider's second flight will improve the analysis of the combined \Spider data set~\cite{spider_bmode_2021,spider1_foregrounds} and benefit the larger community exploring the CMB and modeling Galactic dust.
In the following sections, we briefly summarize the \spider-2 instrument and 280~GHz receiver design, describe the final optical filtering configuration selected for flight, and discuss the receivers' in-flight optical loading and noise performance.

\begin{figure}[b]
\centering
\includegraphics[width=\textwidth]{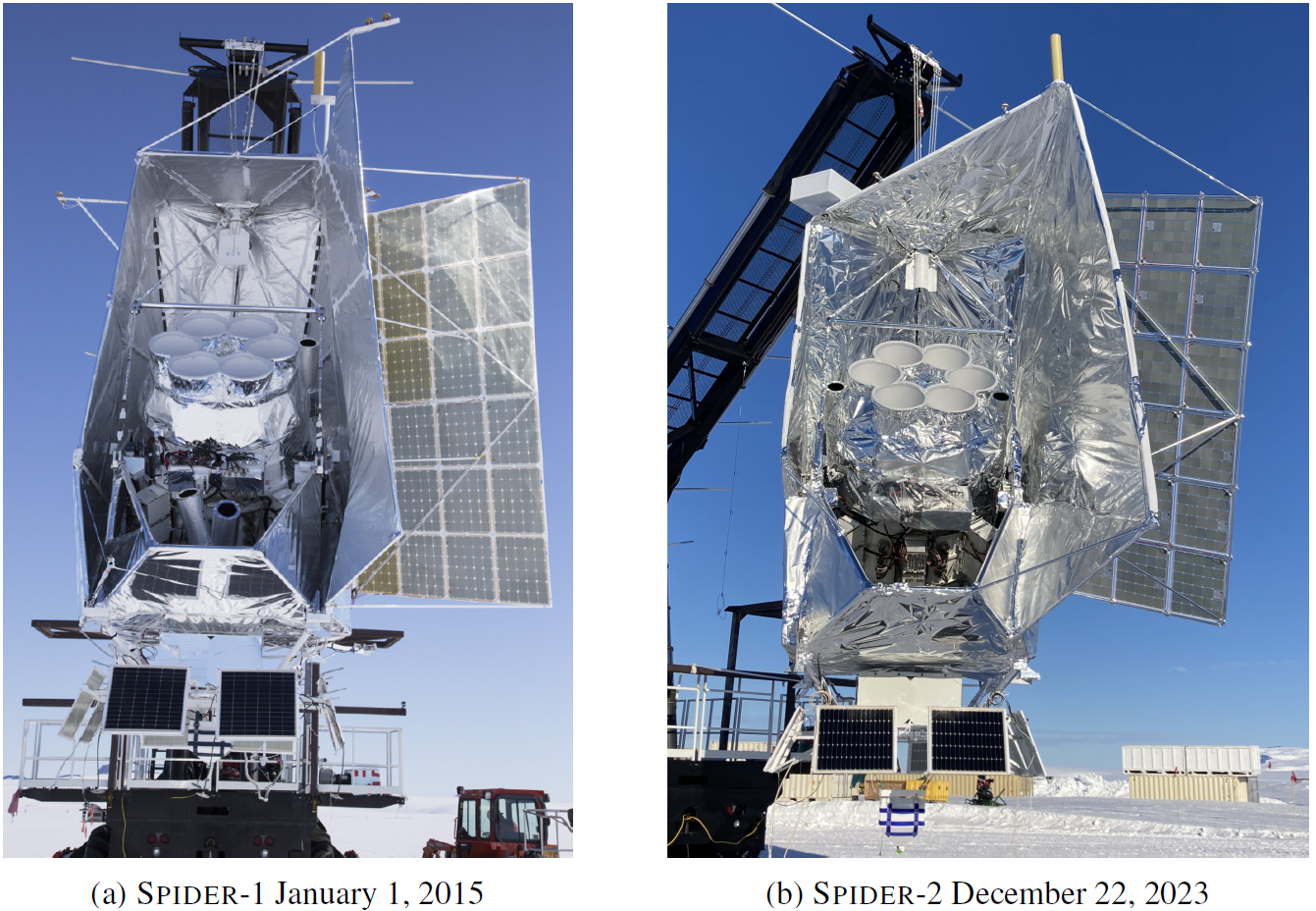}
\caption{Photos of the fully integrated \spider-1 and \spider-2 payloads before their respective launches. Some key components visible in the photographs are the solar panels, star cameras, electronics, telescope forebaffles, and the sunshield.
}
\label{fig:Spiders_Flight_Ready}
\end{figure}

\section{Spider-2 Instrument}
\label{sec:spider2instrument}
\subsection{Payload}
The \spider-2 instrument builds on the heritage of \spider-1, using a similar design and reusing much of the same electronics and hardware -- the visual similarities are evident in pre-flight photos shown in~\cref{fig:Spiders_Flight_Ready}.
The design choices of the \spider-1 instrument and receivers have been well described in several previous publications (Refs.~\citenum{Runyan_2010, Rahlin_2014, Rahlin_thesis, Nagy_thesis, Tucker_thesis}).
For both payloads, a massive $\sim$1300 L liquid helium cryostat houses the six receivers.
Inside the cryostat, the main tank provides 4~K cooling power to the telescope receivers and the cryogenic half-wave plates (HWPs).
Closed-cycle helium-3 sorption refrigerators inside each receiver cool the focal planes and detector assemblies to their 300~mK operating temperatures.
A 20~L superfluid tank (SFT) is filled continuously from the main tank through capillaries, and it cools the telescopes' internal optical baffling and sub-Kelvin refrigerators to 1.6~K.
Helium vapor from the main tank passes through heat exchangers on two stages of vapor-cooled shields, VCS1 and VCS2, cooling the shields and a series of filters mounted at each stage.
In-flight VCS1 hovers around 35~K and VCS2 around 110~K.

The cryostat rests on a lightweight carbon fiber and aluminum gondola surrounded by an aluminized Mylar sunshield.
Forebaffles extending from the telescope apertures shield the detectors from glinting off the Earth's surface and from polarized reflections from the balloon~\cite{Nagy_thesis}.
A solar panel array provides 2~kW of continuous power to the payload.
The instrument is scanned in elevation using dual linear actuators and is scanned in azimuth using a reaction wheel and pivot motor.
The gondola features pin-hole sun sensors, an upgraded gyroscope, a magnetometer, and a dual-boresight star camera system to provide in-flight pointing control and post-flight pointing reconstruction.
The gondola frame, sunshield, and solar panels were entirely rebuilt for the second flight, with several changes to the overall design to reduce mass and improve strength.
Details on the changes and updates to the \spider-2 gondola and sunshield can be found in Refs.~\citenum{Leung_thesis, StevenLi_thesis}. 
A new flight cryostat, based on the previous design\cite{Gudmundsson_2015}, was manufactured at Meyer Tool\cite{meyertool}, and included a larger liquid helium tank that demonstrated improved vacuum performance pre-flight relative to the original unit.
Refs.~\citenum{Song_thesis,Bergman_thesis} report on the integration and pre-flight testing of the \spider-2 cryostat.

\subsection{280 GHz Receivers}
\begin{figure}[!ht]
\centering
\includegraphics[height=13cm]{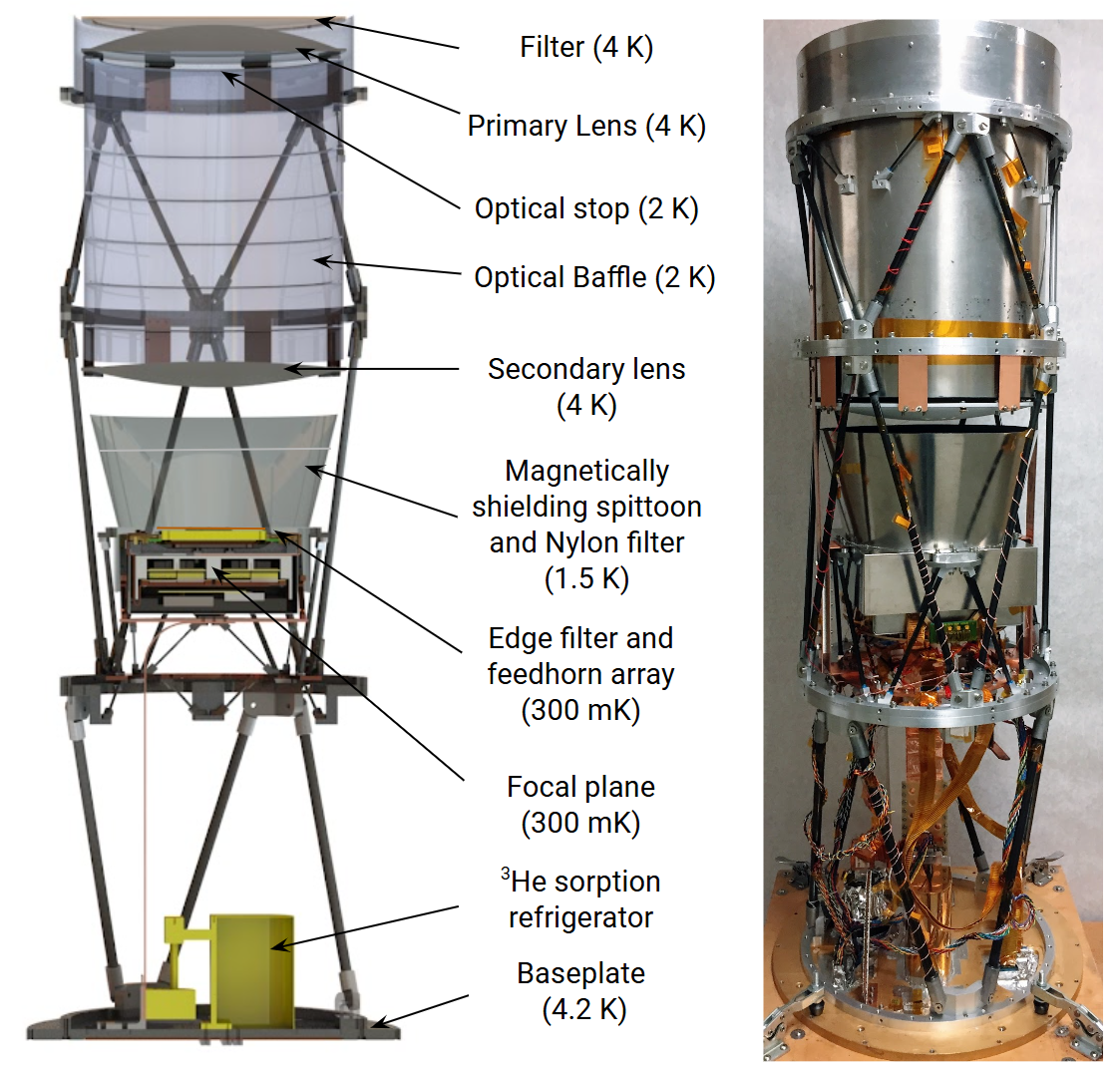}% A
\caption{(Left): A simplified cutaway CAD model of the \textsc{Spider}-2 280~GHz receivers, including the optics, cold baffling systems, filters, focal plane, and $^3$He sorption refrigerator. The rest of the optical stack skyward of the receivers, including the half-wave plate, filters, and vacuum window, are not pictured in the diagram.
(Right): A fully integrated 280~GHz receiver. 
Before installation into a cryostat, the telescope is wrapped in copper-clad G-10 and taped at the seams to be light-tight. 
The 280~GHz receivers stand approximately 1.2~m ($\sim$ 47 in) tall when measured from the top of the base plate to the top of the 4~K filter stack.
This figure was adapted with permission from Shaw et al\cite{Shaw_2020}.}
\label{fig:280_insert}
\end{figure}

The \spider receivers are small-aperture, two-lens, refracting telescopes designed with an emphasis on reducing the internal loading, or the thermal emission, of the inside of the instrument.
\cref{fig:280_insert} shows a photograph and a cutaway drawing of a 280~GHz receiver (the optical chain elements skyward of the receivers are not pictured).
\Spider's three 280~GHz telescopes were constructed from decommissioned 95 and 150~GHz receivers and are outfitted with new focal planes and optics.
The 280~GHz receiver design and pre-flight characterization were described in previous SPIE proceedings (Ref.~\citenum{Shaw_2020}), with more details included in Ref.~\citenum{Shaw_thesis}. 
The following subsections provide a brief overview, and we comment on changes made to the optical stacks since Shaw et al. 2020~\cite{Shaw_2020}.

\subsubsection{Detectors}
The 280~GHz detector arrays are feedhorn-coupled transition-edge sensor (TES) arrays developed and fabricated at NIST~\cite{Hubmayr_2016}.
The detector arrays and sub-Kelvin readout electronics are housed inside each receiver's focal plane unit (FPU) and operate at 300~mK.
\spider's TES bolometers are tailored to take advantage of extremely low background loading in the space-like flight environment, and a TES designed for flight conditions easily saturates during lab testing.
Every \spider bolometer has two TESs with different superconducting critical temperatures, $T_c$, connected in series to address this issue.
An aluminum manganese (AlMn) TES ($T_c\sim450$~mK) provides low noise for science data acquisitions on the 280~GHz detectors, and an aluminum TES with higher $T_c$ ($\sim1.6$~K) extends the bolometer dynamic range for the higher optical loading conditions of lab-based measurements.
Shunt resistors with $\sim1/10$ of the science TESs operational resistance are connected in parallel with the TESs, providing a strong voltage bias.
The TESs are read out using a three-stage time-division multiplexed Superconducting QUantum Interference Device (SQUID) system designed at NIST~\cite{SQUID_multiplexing_2003}.
Multi-Channel Electronics (MCE) developed at the University of British Columbia control the SQUID readout~\cite{SQUID}.
When the 280~GHz receivers are referenced individually, they are called by their FPU naming convention: ``Y3'', ``Y4b'', and ``Y5''.

\subsubsection{Receiver Optics and Filtering}
\label{sec:OpticsandFiltering}
The detectors and focal plane array are located at the center of the receivers, which mount to the base of the main (4~K) helium tank and extend up through cylindrical ports.
The two high-density polyethylene lenses are cooled to 4~K to reduce in-band loading from dielectric loss in the plastic.
An optical baffle surrounding the optical path between the lenses, and a magnetic shield surrounding the focal plane are cooled to 1.6~K.
The optical baffle is blackened to absorb sidelobe power falling outside the optical stop (270~mm diameter), which sits at the top of the baffle \cite{Runyan_2010, Rahlin_2014}.

The rotational symmetry of the telescopes is beneficial for reducing polarized instrumental systematics that could show up in the data as false sky signals.
A half-wave plate (HWP) provides polarization modulation to mitigate beam systematics and increase polarization coverage.
The 280~GHz HWPs are 1.66~mm thick birefringent sapphires from Rubicon Technology\cite{rubicontech} with a $\sim$0.125~mm Cirlex AR coat\cite{Bryan_2010,Bryan_2016}.
The HWPs are mounted to the main helium tank directly skyward of the receivers and are rotated twice per sidereal day in multiples of 22.5 degrees.

We utilize a combination of carefully positioned filters to reflect out-of-band radiation while minimizing the in-band thermal radiation of the filters themselves.
The optical stack includes thin reflective metal mesh shaders~\cite{Tucker_ThermalFiltering}, multi-layered hot-pressed metal-mesh low-pass filters~\cite{metal_mesh_review}, and thicker dielectric filters that shape the desired bandpass. 
The original optics and filtering scheme for the 280~GHz receivers was detailed in Ref.~\citenum{Shaw_2020}.
In that scheme, two reflective IR ``shaders'' mounted directly below the 1/16'' thick ultra-high molecular weight polyethylene vacuum window provide the first defense against infrared radiation. 
Additional shaders are mounted at both VCS stages with two metal-mesh low-pass filters mounted beneath them on the $\sim$35~K  VCS stage.
Another 12~cm$^{-1}$ (360~GHz) cutoff filter is mounted at 4~K just skyward of the primary lens and underneath the sapphire HWP.
A 3/32'' AR-coated nylon filter is located in the 1.6~K magnetic shield spittoon aperture to absorb additional infrared radiation~\cite{Halpern_farIRtransmission}.
Finally, the high-frequency edge of the 280~GHz bandpass is defined by a 10.5~cm$^{-1}$ ($\sim315$~GHz) low-pass metal-mesh filter mounted at 300~mK, directly above the feedhorn array.

\begin{table}[t]
\caption{Optical filter configuration summary for the three 280 GHz receivers with the approximate temperature of the mounting stages. 
Filters prefixed with `C' are reflective IR shaders~\cite{Tucker_ThermalFiltering}. 
The low-pass edge filters~\cite{metal_mesh_review} are denoted by their cutoff frequency in inverse centimeters and are not AR-coated. 
}
\label{tab:Filters}
\begin{center}       
\begin{tabular}{cccc} 
\toprule
\midrule
\rule[-1ex]{0pt}{4ex}   Temperature Stage  &   Y3    &    Y4b   &    Y5      \\  
\midrule
%\rule[-1ex]{0pt}{4ex}     &   1/16"   &    1/16"  &   1/16"   \\ 
\rule[-1ex]{0pt}{4ex}   245 K  &    C15, C30   &    C15, C30   &  C15, C30\\ 
\rule[-1ex]{0pt}{4ex}  110 K    &    C15 $\times$ 3 &   C15 $\times$ 3 & C15 $\times$ 3    \\ 
                     &C30 $\times$ 4  &  C30 $\times$ 4&  C30 $\times$ 4     \\
\rule[-1ex]{0pt}{4ex} 30 K  & Sapphire Filter & Sapphire Filter &  Sapphire Filter  \\
                         &18 cm$^{-1}$&15 cm$^{-1}$&15 cm$^{-1}$\\ 
                        &12 cm$^{-1}$&12 cm$^{-1}$&12 cm$^{-1}$\\ 
                                        
\rule[-1ex]{0pt}{4ex}    8 K  &  HWP  & HWP &  HWP    \\ 
\rule[-1ex]{0pt}{4ex} 4 K  &12 cm$^{-1}$&12 cm$^{-1}$ &12 cm$^{-1}$\\
\rule[-1ex]{0pt}{4ex}  1.6 K & 3/32'' Nylon 6/6 & 3/32'' Nylon 6/6 & 3/32'' Nylon 6/6 \\ 
                              & NDF  &   \textbf{ -- }  & \textbf{--}    \\                    
\rule[-1ex]{0pt}{4ex} 300 mK  &10.5 cm$^{-1}$   &10.5 cm$^{-1}$ &10.5 cm$^{-1}$   \\

\bottomrule
\end{tabular}
\end{center}
\end{table}

\begin{figure}[b]
    \centering
    \includegraphics[width=.65\linewidth]{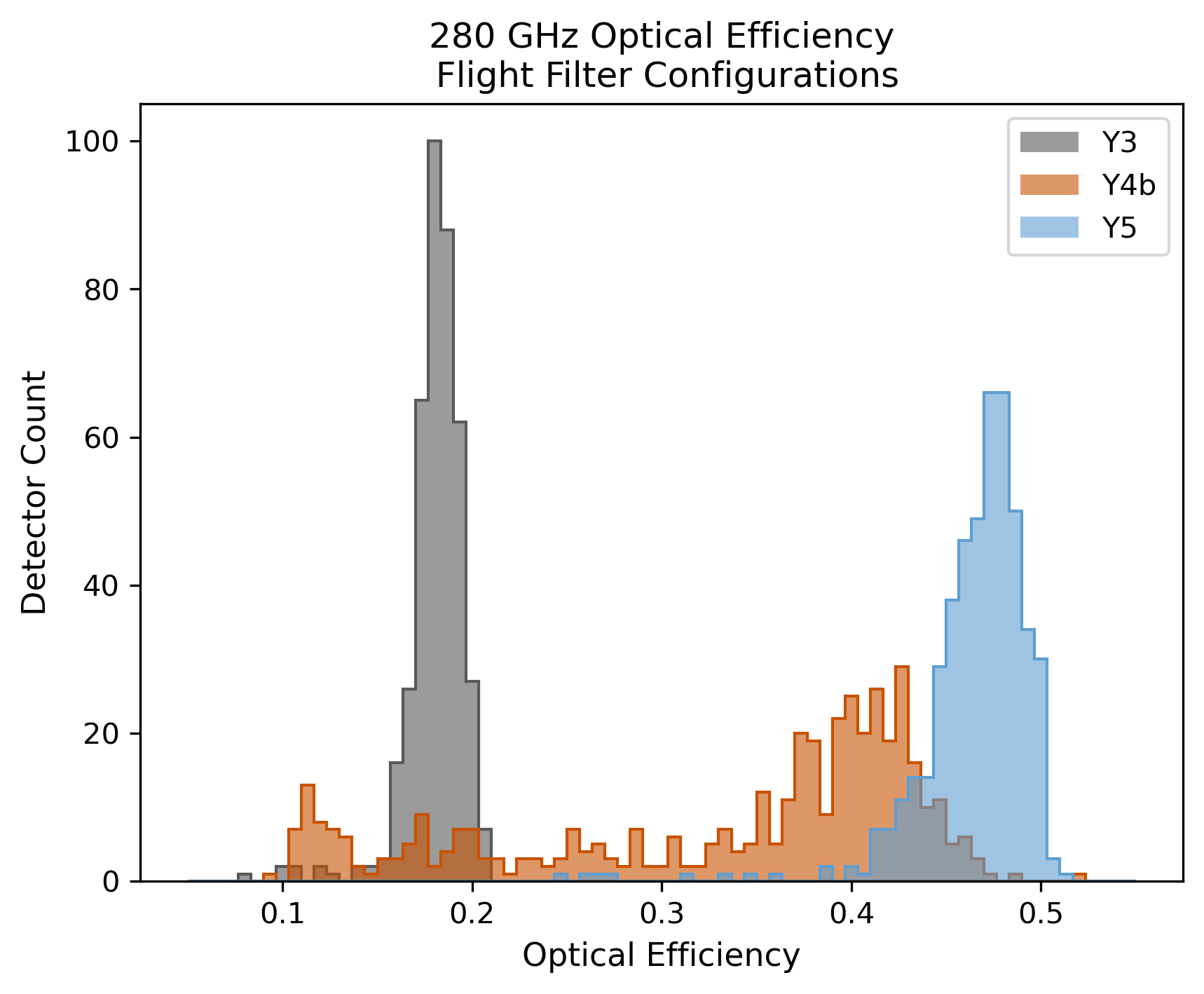}
    \caption{Histogram of 280 GHz detector optical efficiencies with the flight filter configuration detailed in \cref{tab:Filters}. The tail of lower efficiency detectors on Y4b is due to defects in the feedhorn array, and the low efficiency of Y3 is due to an additional NDF.
    See \cref{sec:OpticsandFiltering} for measurement details.
    }
    \label{fig:OptEff_Flight_match}
\end{figure} 
Before flight, we had concerns about high internal loading, but we were limited in our ability to measure it under comparable flight-loading environments.
Laboratory tests suggested that the metal-mesh low-pass filters at the 35~K VCS stage were running warmer than desired, and we worried that their emissivity contributed to high loading measurements.
We made precautionary changes to the optical stacks to balance a reduction in loading while conserving optical efficiency to the greatest possible extent.
The low in-band absorption loss and high thermal conductivity of sapphire are ideal for an IR filter.
So, we moved all IR shaders to the warmer VCS stage and installed sapphire filters in front of the metal mesh filters. 
The intention was that the sapphire filters, similar to an alumina filter, would conduct the IR power load onto the VCS shields rather than re-radiate onto the emissive (and less conductive) filters behind.
The sapphire filters decreased detector responsivity by less than 5\%.
Due to pre-deployment time constraints and what was commercially available immediately, the additional sapphire filters are the same (birefringent) material as the 280~GHz HWPs.
The effect of the birefringent filters on detector polarization angles is captured during pre-flight detector angle measurements using a near-field polarized thermal source~\cite{Nagy_thesis}.

To ensure we would have at least one fully functional 280 GHz receiver, we added a neutral density filter (NDF) to the Y3 receiver, which has the detectors with the lowest saturation power. 
The selected NDF was constructed from four laminated 0.03" sheets of carbon black-filled polyethylene and was measured to have $\sim$42\% transmission.
The NDF was AR-coated and installed beneath the AR-coated Nylon filter at 1.6~K.
A comprehensive list of the filters installed in each of the 280 GHz receivers is shown in~\cref{tab:Filters}.

During pre-flight characterization tests, we measured the optical responsivity, $dP/dT$, of each receiver using an aperture filling thermal load at room temperature and 77~K. 
Load (IV) curves acquired on the superconducting transition of the higher-$T_c$ aluminum TESs enable the instrument characterization without temporary neutral density filters that would affect the interpretation of the measurements.
The detector optical efficiencies are derived assuming an ideal detector efficiency of 0.871~pW/\krj (a top hat bandwidth of 63~GHz).
The 280~GHz detector optical efficiencies measured before flight are presented in~\cref{fig:OptEff_Flight_match}.
With the additional sapphire filter, the bulk of Y5's detectors are around 47\% efficient, and the higher optical efficiency detectors on Y4b cluster around 40\%.
The additional NDF on Y3 reduces the optical efficiency to $\sim$18\%.
For a more detailed discussion of pre-flight characterization methods and the filtering configuration chosen for the flight, please refer to Ref.~\citenum{Shaw_thesis}.

\begin{figure}[!h]
    \centering
    \includegraphics[width=\textwidth]{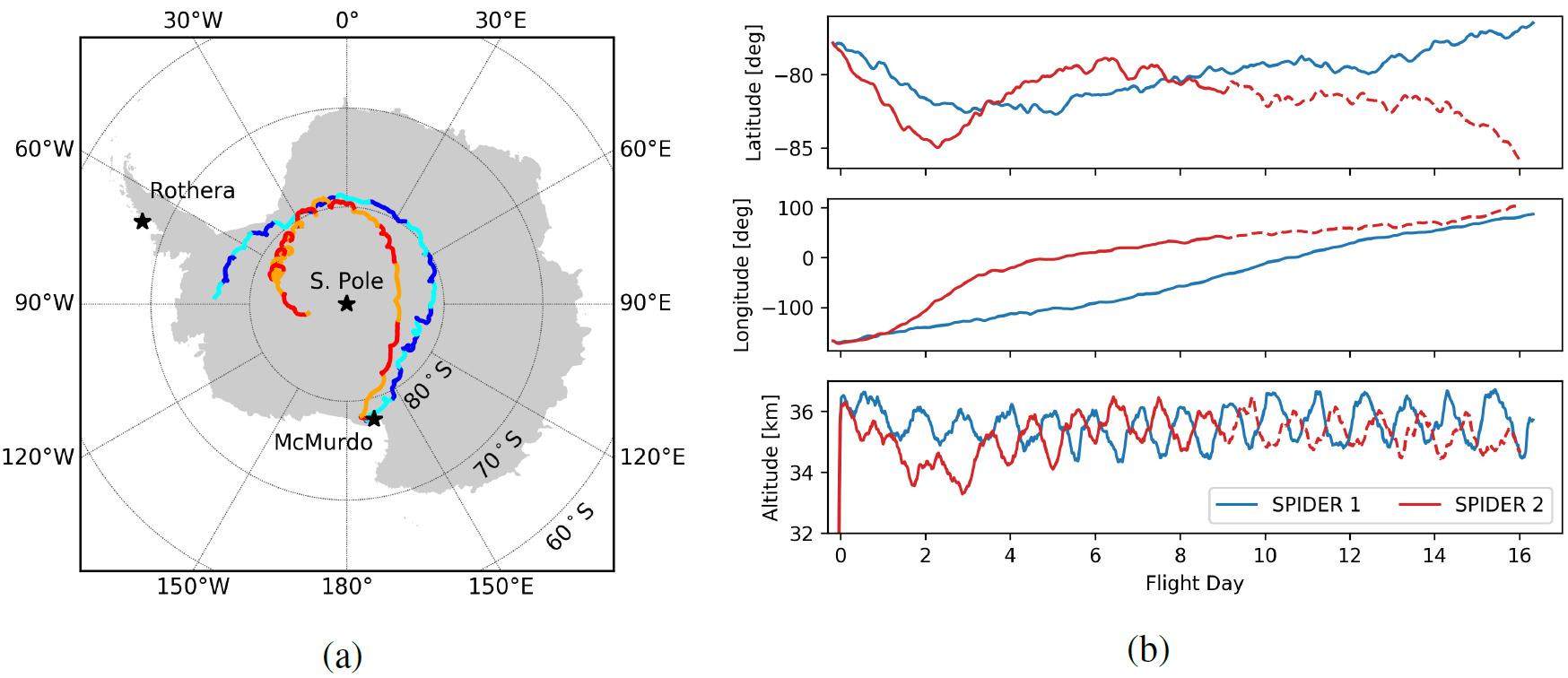}
    \vspace{1mm}
    \caption{(a): Flight trajectory during \spider's flights. \spider-1 in blue/cyan, and \spider-2 in red/orange. 
    Alternating colors along each track denote days. (b): Payload longitude (top), latitude (middle), and altitude (bottom) during \spider's flights. T=0 marks when the payloads reached an altitude of 36~km. The solid lines on \spider-2 data correspond to the nominal science observation period.}
    \label{fig:FlightDetails}
\end{figure}

\section{\Spider-2 Flight}
The \spider-2 payload launched on December 22, 2022, at 10:26 AM NZDT.
The payload circumnavigated the continent at an average altitude of 35~km for 16 days.
The payload hovered over the Transantarctic Mountain range for several days and was terminated on January 7, 2023, near Hercules Dome, Antarctica. 
\cref{fig:FlightDetails}a shows the payload flight path over the Antarctic continent alongside the \spider-1 flight path in 2015.
Both flights' longitude, latitude, and altitude profiles are shown in~\cref{fig:FlightDetails}b. 
Within three weeks of flight termination, the data drives and payload hardware were fully recovered with assistance from Kenn Borek Air and South Pole Station personnel.

The majority of the payload subsystems operated as designed throughout the flight.
All power, pointing, thermal management, detector monitoring, and half-wave plate rotation systems operated successfully.
Additionally, all of \spider's receivers performed exceptionally well, as discussed below. 
The flight nonetheless suffered several challenges, partially due to the tight schedule and logistical constraints of the 2022-23 season.
We were not able to fill the liquid helium tank to full capacity before launch, so science data collection ceased on the ninth day of flight after the tanks emptied. 
We also found evidence after the launch of a helium leak into the vacuum space (similar to \spider-1~\cite{Gudmundsson_2015}).
Finally, the payload's star cameras did not function at float, so we relied on the remaining pointing sensors for attitude control. 
Given \spider's relatively modest pointing requirements, however, we expect to be able to adequately reconstruct a pointing solution using the other pointing sensors on the gondola (sun sensors, magnetometer, gyroscopes) and via cross-correlation between \planck temperature maps and \spider's own data.
Significant progress has been made on this iterative process toward science-quality maps.

\section{In-Flight Optical Loading}
\label{sec:inflight_loading}

\begin{figure}[b]
    \centering    
    \includegraphics[width=.65\textwidth]{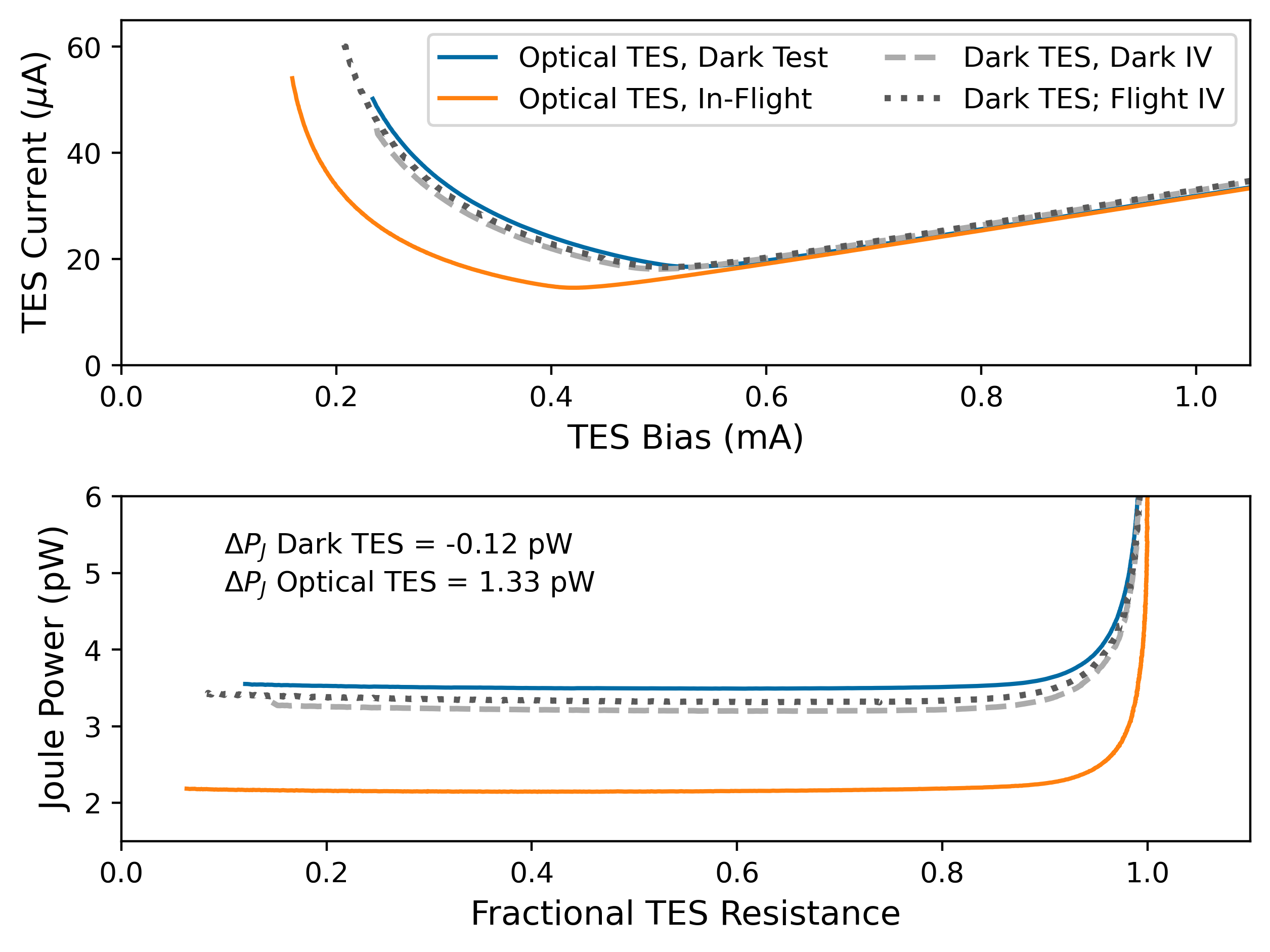}
    \caption{Representative IV curves collected during dark testing and flight for an optical TES and a non-optical (dark) TES on the Y4b array.
    The top panel shows the TES current versus the current bias, and the bottom panel shows TES power versus fractional TES resistance. 
    The FPU temperature was 333~mK and 326~mK for the dark and in-flight IV curves, respectively.
    The raw difference in Joule power for each TES is printed in the second panel. 
    For the optical TES ($\sim 40$\% efficiency), the difference is a combination of effects from optical input and lower focal plane temperature. 
    The small but non-zero $\Delta P_J$ for the dark TES is consistent with the expected effect of the difference in focal plane temperatures.
    }
    
    \label{fig:DarkandFlightIVs}
\end{figure}

\begin{figure}[tb]
    \centering
    \includegraphics[width=\linewidth]{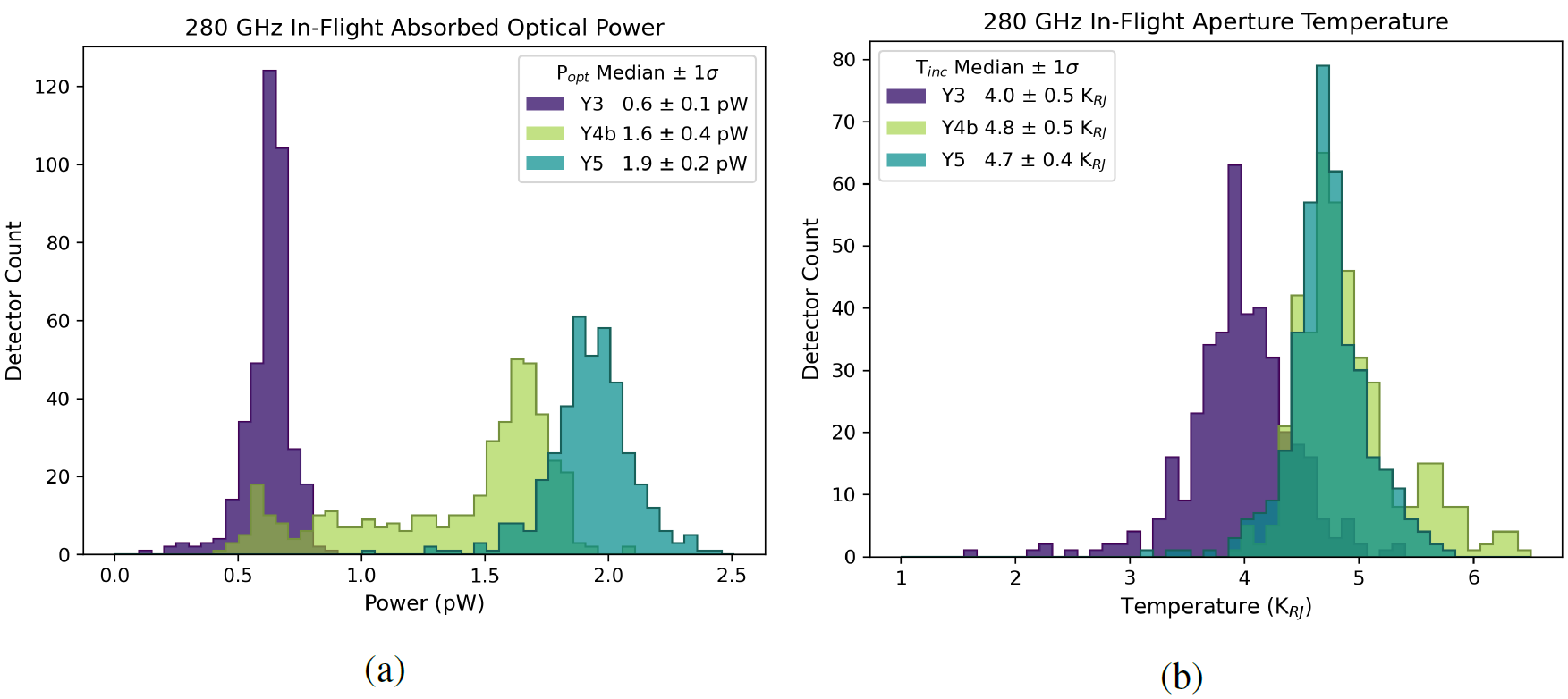}    
    \caption{In-flight loading averages at 280~GHz with detector data grouped by FPU. (a): Absorbed optical power (b): Apparent aperture temperature, including sky and internal loading, calibrated using detector responsivity.}
    \label{fig:280_Flight_Loading}
\end{figure}

The radiative power absorbed by the detectors in flight (``optical loading'') is a critical performance parameter due to its effect on instrumental noise.
We extract a measurement of the optical loading on the detectors in the near-space environment by comparing IV curves taken at float to those taken during dark testing. 
After correcting for FPU temperature variations, the difference of Joule power on the superconducting transition of the in-flight IV curve and the saturation power from dark testing yields direct estimates of the optical power absorbed by each detector.
During dark testing, a gold-plated silicon mirror placed on the feedhorn arrays ($\sim$300~mK) isolates the thermal and electrical TES properties from optical input. 
We measure the detector thermal and electrical parameters (\eg, $G$, $T_c$, $\beta$, $P_{sat}(T_c)$, $R_{shunt}$) and use them to correct for saturation power differences due to FPU temperature variation and for comparison with optical datasets.
\cref{fig:DarkandFlightIVs} shows representative IV curves from flight and dark testing for a dark TES and a single optical TES to demonstrate optical loading measurement.
The optical loading values reported in this paper have been corrected for FPU temperature variation using the dark detector properties and methods outlined in Refs.~\citenum{Shaw_thesis,Bergman_thesis,Shaw_2020}.

The distribution of in-flight absorbed power, $P_{opt}$, for all 280~GHz detectors is shown in~\cref{fig:280_Flight_Loading}a.
The loading reported for each detector is the average of 4--5 load curves taken in the first 24 hours at float.
The median absorbed power measured on Y3, Y4b, and Y5 is 0.6, 1.6, and 1.9~pW, respectively.
These powers are well clear of saturation for all detectors; the absorbed powers on Y5 detectors are just above half their saturation powers, while Y3 and Y4b are less.
For \spider's 280~GHz receivers, the CMB and atmosphere are only a small contribution to the total absorbed power; even for Y5, the 280~GHz receiver with the highest optical efficiency, their combined contribution is expected to be less than 0.12~pW~\cite{Bergman_thesis}.

The power absorbed by each detector heavily depends on its optical efficiency (\cref{fig:OptEff_Flight_match}).
We can alternatively express the loading as an apparent aperture temperature, $T_{inc}$, by dividing $P_{opt}$ by the detector responsivity. 
$T_{inc}$ represents the in-band thermal load of everything skyward of the focal plane (optics, filters, window, baffles, atmosphere, and CMB), treated as if it originated at the aperture.
Histograms of $T_{inc}$ are presented in~\cref{fig:280_Flight_Loading}b.
The aperture temperatures seen by Y4b and Y5 are consistent with $\sim$4.8~\krj; this behavior was anticipated because Y4b and Y5 have nominally identical filter stacks.
The loading measured by Y3 is closer to 3.9~\krj, presumably reduced by the effects of the NDF installed at the 1.6~K stage.

We can further explore to what extent detector loading varies throughout the flight.
For \spider-1, loading variations were primarily associated with fluctuations in the FPU temperatures between fridge cycles and were on the order of 10\% of the total optical loading of $\sim$0.3~pW at 150~GHz~\cite{Gambrel_thesis}.
\spider-2 saw substantially larger temperature fluctuations in the temperatures of its cryogenic stages and associated optical elements. 
Thermometer readings on the first $\sim$35~K VCS stage varied between 20 and 45~K, primarily due to hysteresis in the helium boil-off flow from the main LHe tank vent. 
The thermometers on the second VCS stage ($\sim$110~K) showed a $\sim$5~K oscillation. 
A thermistor attached to the window clamp on the Y4b aperture dropped from 270~K to 240~K in under 24 hours after reaching float altitude, then subsequently oscillated between 240--245~K during the diurnal cycles (This behavior is not necessarily unlike \spider-1).
We must thus consider the loading variations that might result from these temperature changes.

To estimate the loading variations, we compare load curves taken on Y4b in the first 24 hours of flight to one taken later in the flight, shortly after a fridge cycle when the VCS temperatures are at a minimum.
The difference in absorbed power is near 0.15~pW for the higher optical efficiency detectors on Y4b ($\Delta T_{inc}\sim$0.4~\krj).
With 0.15~pW as a lower limit on the range of loading fluctuations for the higher optical efficiency detectors, we expect the photon noise levels to fluctuate by 5\% over the temperature oscillation period, representing a \sims5\% cost in mapping speed. 

\section{Noise Performance}
In this section, we present initial studies of the white noise performance of the 280~GHz detectors throughout \spider's second flight.
Note that the analysis of this data set is still in its early stages: data cleaning is still being finalized, and sky maps are not yet mature enough to calibrate detector data against sky signals.
We can nonetheless make initial estimates of noise performance, focusing in particular on periods in which the payload is not scanning, and give an approximate calibration to on-sky sensitivity based upon laboratory calibration measurements and load curves acquired during flight.

\subsection{Calibration}
\label{subsec:noisecalib}
We calibrate raw detector time streams to current, $I$, based on a linearized model of TES response.
The calibrated TES $I$ is derived using measurements of the detector shunt resistance, readout wire resistance, and other readout parameters.
We can assess the calibration from the superconducting slope of calibrated IV curves; the superconducting slope should equal unity for a well-calibrated current.
From inspections of IV curves, we expect the error due to our current calibration to be less than 1.5\% for all detectors.

To express detector noise in units of equivalent absorbed power (the noise equivalent power, or NEP), we must calibrate time-ordered data from current to power. 
We approximate the TES responsivity, $s_I$, defined as the change in TES current for a small change in optical power dissipated on the bolometer island, using the responsivity equation derived in Irwin \& Hilton (2005), evaluated at the zero-frequency limit~\cite{Irwin2005Hilton},
\begin{equation}
    s_I(\omega=0) = -\frac{1}{IR}\left(\frac{L}{\tau_{el}R\mathcal{L}_I}+\left(1 - \frac{R_{sh}}{R}\right)\right)^{-1},
\end{equation}
where $R$ and $I$ are the TES resistance and current, $R_{sh}$ is a shunt resistor connected in parallel to the TES,  $L$ is the TES loop inductance,  $\mathcal{L}_I$ is the TES loop gain, and $\tau_{el}$ is the TES electrical time constant.
In the limit of high loop gain, the calibration factor we require reduces to the following:
\begin{equation}
     \frac{dP}{dI} = s_I^{-1}(\omega=0) \approx -I\left(R - R_{sh}\right). \label{eq:dPdI_conversion}
\end{equation}
We use \cref{eq:dPdI_conversion} to convert from TES current to absorbed power, then inflate the results by 6\% to account for the estimated effect of finite loop gain.
We use load curves acquired in the middle of the stop mode data set to derive values of TES current and resistance at the appropriate voltage bias for each data point.
This method assumes that the optical loading is constant throughout the dataset. 
In practice, we know that the absorbed optical power on the higher optical efficiency detectors can fluctuate by at least 0.15~pW throughout flight due to the variation in cryostat temperatures~\cite{Shaw_thesis}. 
To bracket the error introduced by the choice of reference load curve, we repeated the noise analysis on Y4b, calibrating with a load curve taken later in flight while the cryostat's vapor-cooled shields were at their coldest.
The resulting decrease in mean Y4b detector NEP is less than 0.2\%, suggesting this calibration error is negligible.

Finally, we use laboratory measurements of optical response to calibrate from NEP to NET$_{\text{CMB}}$, or noise-equivalent temperature, representing the detector sensitivity to the CMB signal.
To calculate NET we divide NEP by $dP/dT_{CMB}$, computed in the small-bandwidth approximation as follows:
\begin{align}
   \frac{dP}{dT_{CMB}} &= \frac{d}{dT_{CMB}}\left(\sqrt{2}\int d\nu\, \eta\, B(\nu,T) \,dA\, d\Omega\right)\\
   &\approx \sqrt{2}\,\frac{dP}{dT} 
                      \frac{x^2e^x}{(e^x - 1)^2},\;\; x = \frac{h\nu_0}{k_B T_{CMB}}.
\end{align}
Here $dP/dT$ is the optical responsivity (in the Rayleigh-Jeans limit) measured in the laboratory for each detector, $\nu_0$ is the band center and the factor of $\sqrt{2}$ is included to convert the NET from units of $\sqrt{Hz}^{-1}$ to $\sqrt{s}$.
We use array-averaged band centers $\nu_0$: 274.7~GHz, 275.0~GHz, and 269.0~GHz for Y3, Y4b, and Y5 respectively. 
The band centers for Y3 and Y5 are the averages of Fourier transform spectroscopy measurements made with a custom Martin-Puplett interferometer~\cite{Song_thesis}. 
The measured band centers showed a $1\sigma$ spread around these medians of 4.5~GHz (2.6~GHz) for Y3 (Y5).
The Y4b bandpass measurement was not completed before flight due to pandemic constraints and logistical challenges before and during Antarctic deployment which forced us to prioritize meeting launch deadlines over conducting the diagnostic bandpass measurement.
Since the lower band center of the Y5 array is believed to be a manufacturing error in the size of the feedhorn waveguides, we chose to use a round value near the measurement of the Y3 array for the Y4b band center in this analysis.
A 5~GHz discrepancy in the true band center would introduce a $\sim$5\% error on the NET values reported below.

\begin{figure}[tbh]
    \centering
    \includegraphics[width=.95\textwidth]{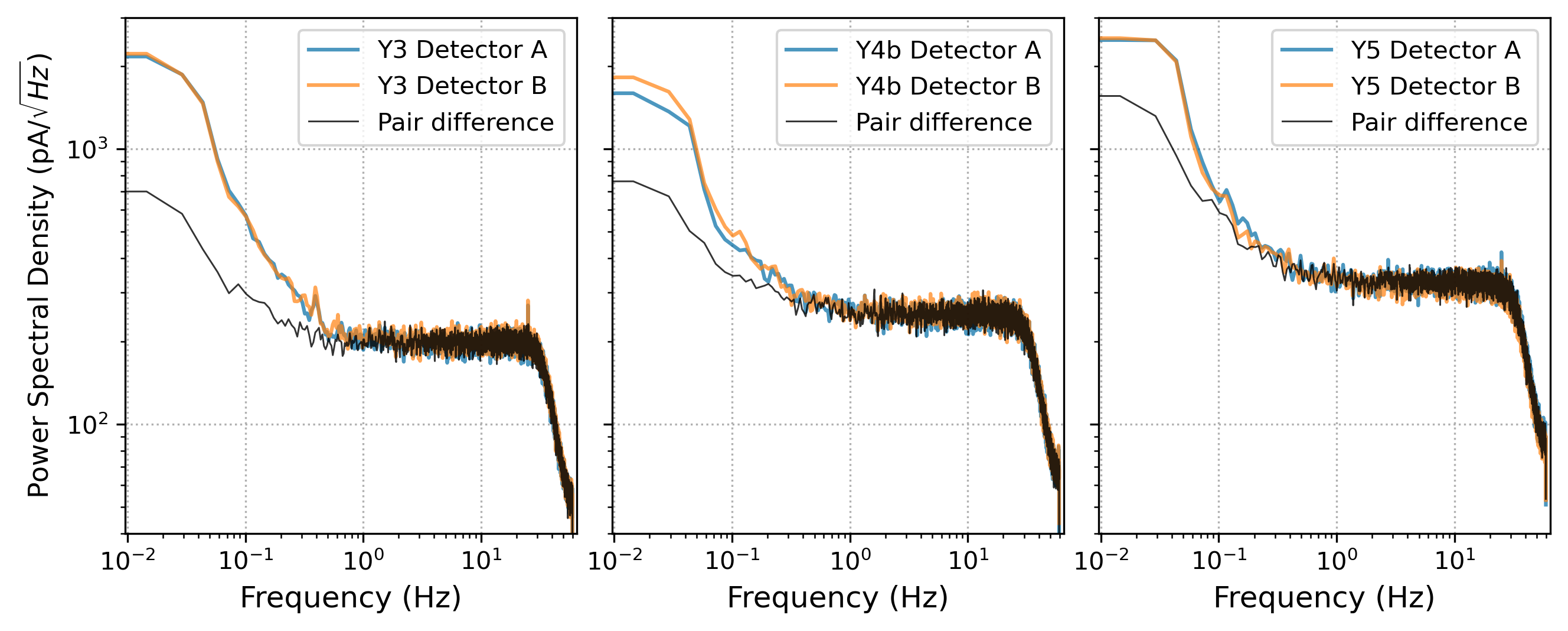}
    \caption{Power spectral density pair-difference plots from representative detector pairs on each FPU.  The PSDs shown are the average of the PSDs calculated at each time chunk. The low-frequency correlated noise in the PSDs of the pair-difference time streams is suppressed.}
    \label{fig:NEI_stopmodepairdiff_PSDs}
\end{figure}

\subsection{Stop Mode Noise}
During the flight, we took several short data sets with the payload in ``stop mode,'' where the pointing controls maintain the payload's attitude in one direction. 
While the payload is in stop mode, the 1/$f$ knee from scan-synchronous noise is suppressed, and the baseline detector performance can be evaluated more easily.
These stop mode data sets were acquired during the first 24 hours at float, between initial system tests that required the antennas to be powered on.
The usable times where the antennas were off amount to a $\sim$115~minute dataset.

\subsubsection{Noise Power Spectral Density}
The time-ordered data (TOD) used in the noise analysis is divided into smaller chunks and cleaned before calculating the detectors' power spectral density (PSD).
We apply flags to remove intermittent antenna pings and abnormal detector performance and stitch together large discontinuities (``flux jumps'').
We divide the full TOD into chunks of $2^{13}$ samples, each $\sim$68 seconds long.
Any chunk with a flagged data fraction higher than 80\% is excluded from the analysis. 
Finally, we remove a linear fit to each chunk, gap-fill the flagged regions with white noise of the same level as surrounding data, and compute the PSD.

\cref{fig:NEI_stopmodepairdiff_PSDs} shows representative noise PSDs from each of the three FPUs, expressed in units of equivalent TES current per square root bandwidth --- the noise equivalent current (NEI).
For each FPU, we plot PSDs for one detector pair (called A and B channels) and their difference ($(A-B)/\sqrt(2)$). 
The plotted PSDs are averages from all stop-mode time chunks. 
The noise level on the 280~GHz bolometers is reasonably flat across frequencies between the 1/$f$ knee ($\sim$0.4~Hz) to the roll-off at 30~Hz from the internal MCE digital filter. 
The majority of detectors have a low level of correlated noise resulting from temperature and loading drifts and some small, unresolved steps in the time streams.
In the pair-difference PSD, the effects of the low-frequency correlated noise are suppressed. 
For the majority of detectors, the white noise levels between 0.5 and 3~Hz also show slight improvement; the median detector NEI decreases by 2.6\%, 4.0\%, and 1.0\% for the Y3, Y4b, and Y5 arrays, respectively. 
A few readout columns exhibited intermittent, highly correlated electrical noise during flight, often presenting as frequent small jumps in the time streams. 
About 6\% of detectors on each array are affected.
The improvement in white noise levels in the pair difference analysis on these columns is more drastic: 24\%, 62\%, and 32\% for Y3, Y4b, and Y5, respectively.
While the pair-difference analysis would reduce correlated noise, we choose to present results from the single-detector noise analysis below.

The estimate of each detector's noise level in the stop mode analysis is derived from the median white noise value across all time chunks. 
We calculate the white noise level during each time chunk from the mean of the PSD between 0.5 and 3~Hz.
This frequency range roughly covers the few-degree angular scale of interest for \spider's beam size and scanning speed.
The typical scanning speed during flight was 2.5~degrees per second, so the degree-scale signal drops off around 2.5~Hz.

\begin{figure}[tb]
    \centering    
    \includegraphics[width=.65\textwidth]{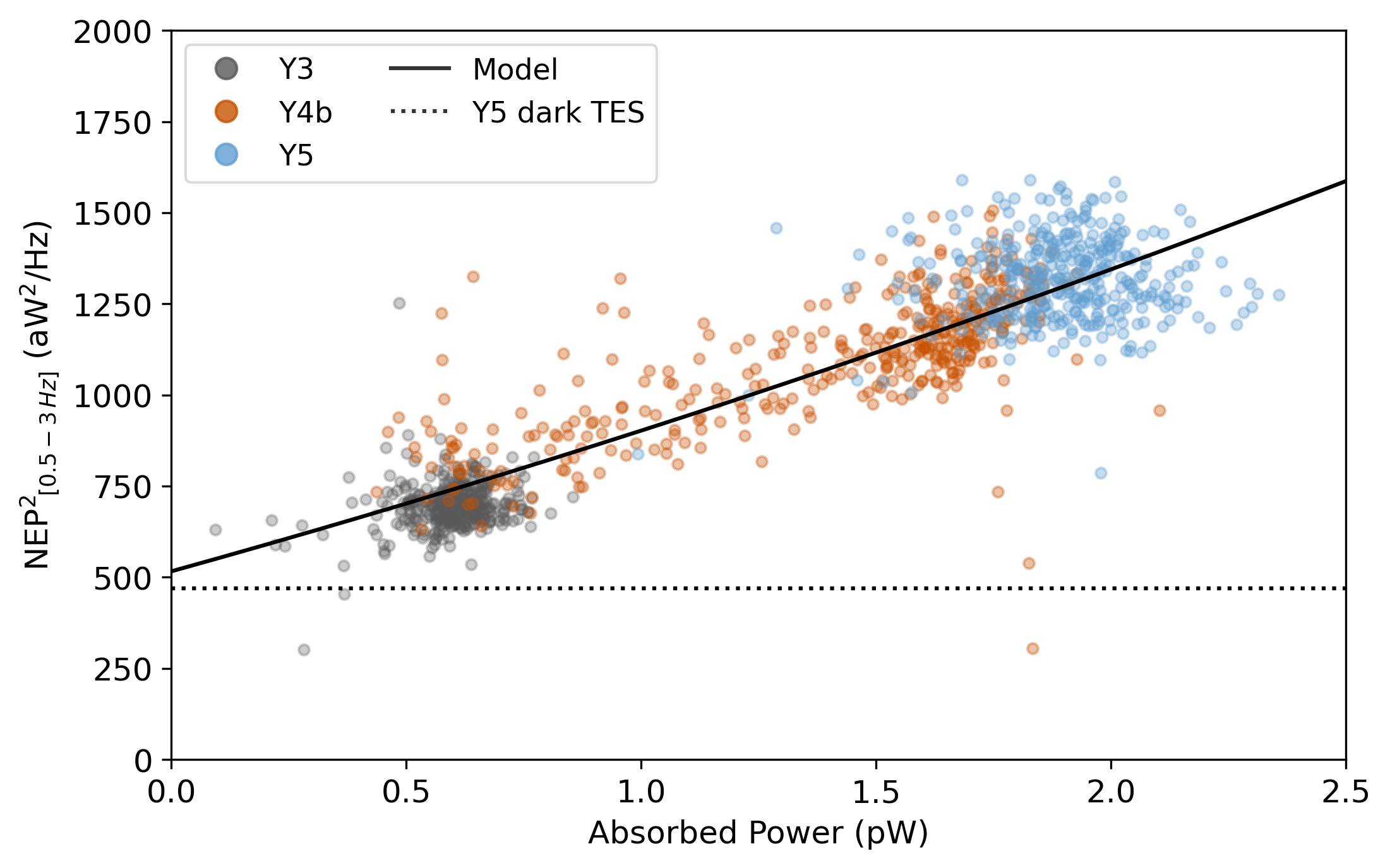}
    \caption{Median measured NEP$^2$ versus absorbed power $P_{opt}$, colored by receiver. 
    The solid black line represents a model of the expected photon noise NEP$^2_\gamma$ per absorbed power for a detector with band center and bandwidth matching the average Y3 and Y5 bandpass ([$\nu_0$, $\Delta\nu$] = [270, 71~GHz]) with an offset fitted to the data to represent other detector noise contributions (e.g., amplifier noise and thermal fluctuation noise).
    The Y5 dark TES noise level is shown for comparison; the dark TES was biased low on its superconducting transition ($\sim$ 0.2~$R_N$) during science data acquisition but nonetheless shows good agreement with the fitted offset.
    Detectors with excess correlated electrical noise have been omitted from the plot for clarity.
    }
    \label{fig:NEPvPabs_280}
\end{figure}

\subsubsection{Noise and Absorbed Optical Power}
The photon noise contribution to total power on a polarized bolometer is expected to scale with absorbed optical power as \begin{equation}
    \text{NEP}^2_\gamma = 2 h \nu_0 P_{opt} + 2 P_{opt}^2/\Delta\nu,
\end{equation}
where $\nu_0$ is the band center and $\Delta\nu$ is the bandpass.
In \spider's regime, the first term is expected to dominate.
The spread in optical efficiencies, and thus the spread in $P_{opt}$, of the 280~GHz detectors allows us to demonstrate the dependence of bolometer noise on absorbed optical power. 

In~\cref{fig:NEPvPabs_280}, we plot the median noise level (NEP$^2$) against absorbed power for each of the 280~GHz detectors. 
The absorbed power is derived using the in-flight load curves (\cref{fig:280_Flight_Loading}a).
We fit the data to a simple model: NEP$^2_\text{Tot}$ = NEP$^2_\gamma$ + offset.
We model NEP$^2_\gamma$ as a detector with the average bandpass measurements of Y3 and Y5 ($\nu_0$=270~GHz, and $\Delta\nu$=71~GHz). 
The fitted offset closely matches the dark TES noise levels on Y5, which should reflect all non-optical noise sources in the bolometers and readout chain, such as amplifier noise and thermal fluctuation noise. 
We attribute any noise in excess of the offset/dark TES noise to photon noise.
For this comparison, a small fraction (6\%) of detectors from columns with highly correlated electrical noise are masked.
The remaining data agree with photon noise-limited performance for the high-efficiency detectors. 

\begin{figure}[tbh]
    \centering
    \includegraphics[width=.9\linewidth]{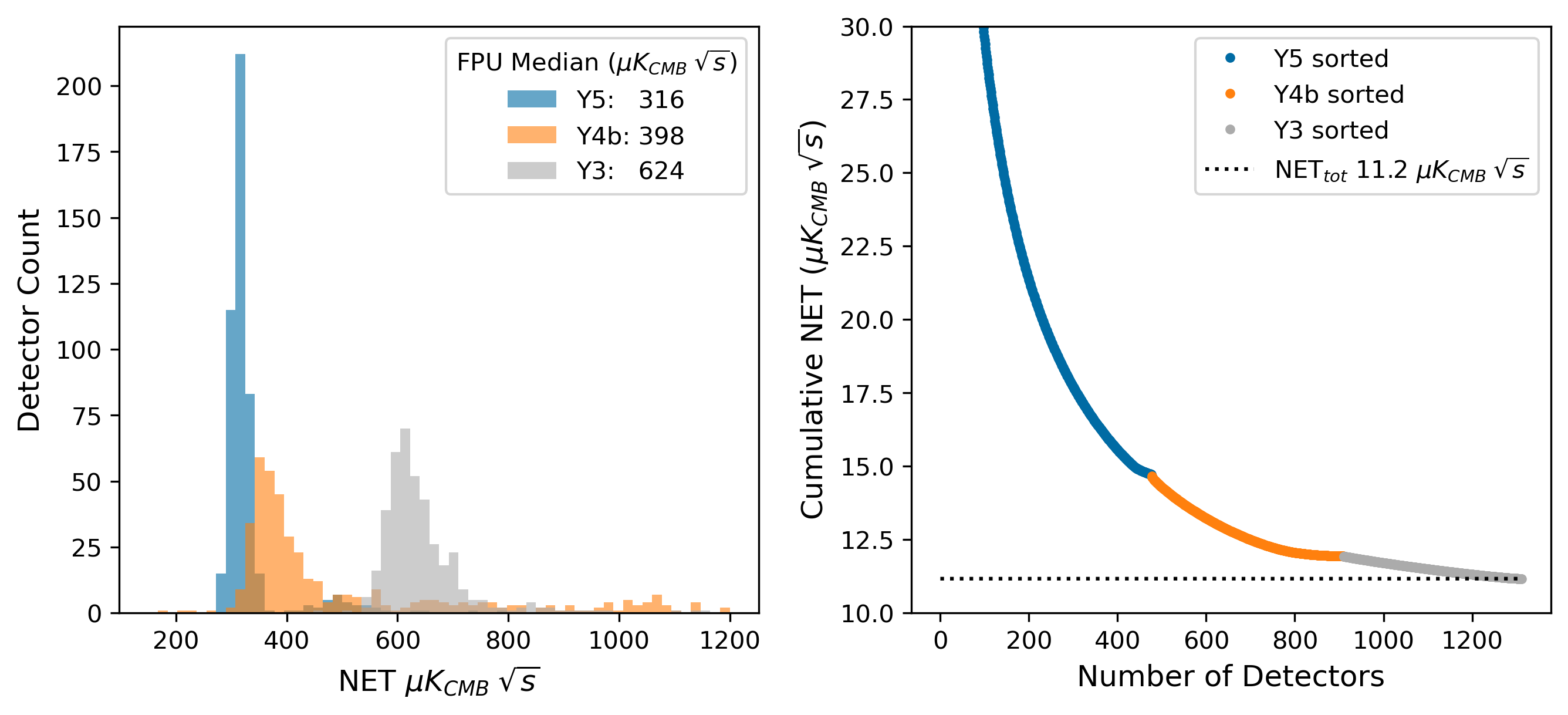}
    \caption{(Left): Histogram of median detector NET colored by focal plane for all stop-mode data.  (Right): Cumulative NET for all 280~GHz detectors. The data are grouped by FPU and sorted from least to most sensitive.
    The overlap of detector sensitivity between FPUs results in cusps in the cumulative NET curve. The black dotted line marks the total sensitivity.}
   \label{fig:CumulativeNET_280GHz}
\end{figure}

\subsubsection{280 GHz Instantaneous Sensitivity}
The left-hand side of ~\cref{fig:CumulativeNET_280GHz} shows histograms of the single-detector median NET, estimated using the calibration described in~\cref{subsec:noisecalib}. 
The achieved sensitivity on Y5 and the high-optical efficiency detectors on Y4b cluster around 350~\mukcmbrts, whereas the extra neutral density filter on Y3 suppresses that array's sensitivity. 
The median detector NET on Y3, Y4b, and Y5 are 624, 398, and 316 \mukcmbrts, respectively; 
considering just the Y4b detectors with optical efficiency above 30\%, that array's median drops to 372~\mukcmbrts.

The right-hand side of \cref{fig:CumulativeNET_280GHz} shows a cumulative plot of the total NET (NET$_{Tot}$) for all 280~GHz detectors. 
The data are grouped by FPU and sorted from least to most sensitive.
The FPU sensitivity for Y5 alone is $\sim$15~\mukcmbrts. 
This exceptional performance results from Y5's high yield, high optical efficiency, and low noise, surpassing the proposal-level forecast of 18~\mukcmbrts (per receiver) in Ref.~\citenum{Fraisse_2013}. 
Although Y4b and Y3 are moderately less sensitive, all three receivers individually exhibit better instantaneous NET than any previously published result above 220~GHz.
The combined 280~GHz sensitivity is \sims11.2~\mukcmbrts, which is on par with the forecast total sensitivity in Fraisse et al. 2013 (\sims10.5~\mukcmbrts)\cite{Fraisse_2013}.

\begin{figure}[tbh]
    \centering
    \includegraphics[width=0.9\linewidth]{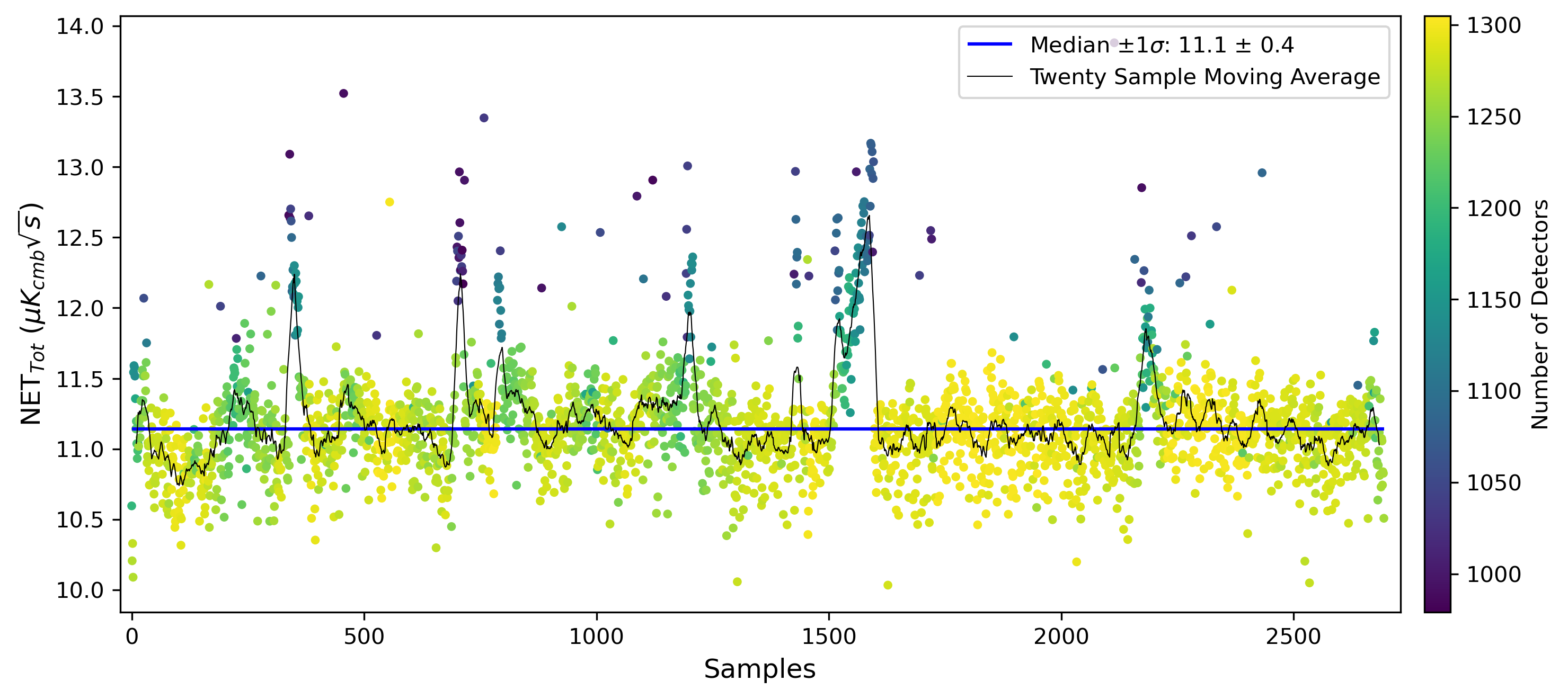}
    \caption{280~GHz $\text{NET}_{\text{Tot}}$ for a very conservatively flagged subset of scanning data during flight. Each data point represents $\sim$1~minute of data and is colored by the number of detectors included in the NET$_\text{Tot}$
calculation. 
The black line is a moving average of the data samples, and the horizontal line marks the median value. 
We find similar baseline noise performance in the scanning data as in the stop mode data.
}
\label{fig:FullFlightCumulativeNET_280GHz}
\end{figure}

\subsection{Full-Flight Scanning Noise}
\label{sec:fullflightnoise}
In this section, we extend the noise analysis to a subset of the science scanning data. 
During the data selection, we subdivide the TOD into $2^{13}$ sample-long chunks ($\sim$68 seconds, approximately one full scan period) and apply very strict and conservative flagging and time stream cleaning.
We exclude time chunks in which more than 25\% of nominally live detectors are flagged.
The remaining data, when combined, is approximately 51~hours long.
To address scan-synchronous signals, we apply a third-order polynomial filter to each chunk and report the average white noise levels between 2 and 4~Hz. 
We then evaluate NET$_{\text{Tot}}$ using all unflagged 280~GHz detectors for each $\sim$1~minute-long chunk.
\cref{fig:FullFlightCumulativeNET_280GHz} shows the behavior in NET$_{\text{Tot}}$ for the 51-hour dataset, where the data is colored by the number of detectors used in the cumulative sum.
The median NET$_{\text{Tot}}$ across the dataset is $11.1\pm0.4$~\mukcmbrts \textemdash similar to the stop-mode dataset.
The larger fluctuations in sensitivity shown in the figure correlate strongly with higher detector flagging fractions from events like fridge cycles or, in some cases, sub-optimal detector biasing. 
One instance of increased NET$_{\text{Tot}}$, around sample 1500, is from a brief interruption in the bias tracking algorithm on the Y5 FPU.
The smaller variations in NET$_{\text{Tot}}$ are likely due to fluctuations in the cryogenic temperatures and optical loading on the TESs, as discussed in~\cref{sec:inflight_loading}. 
The cryogenic temperatures fluctuated during the first half of the scanning mode dataset but were more stable in the second half; the trend in NET$_{\text{Tot}}$ follows a similar pattern.
However, we leave a more detailed review of those features for a future publication.
Overall, the baseline performance demonstrates the stability of the balloon platform for CMB observations.

We emphasize that these are only preliminary estimates on NET.
The absolute calibration for \spider's science analysis will be derived by cross-calibrating degree-scale power in our maps against \planck's maps of the CMB temperature anisotropies~\cite{spider_bmode_2021}.

\section{Conclusion}
The \spider CMB experiment has completed its second and final NASA Long Duration Balloon flight and mapped the Southern skies in three frequency bands: 95, 150, and 280~GHz. 
The three novel 280~GHz receivers deployed during the second flight were the first implementations of the NIST feedhorn-coupled TES array technology at 280~GHz and the first to use the technology for observations from a balloon.
The detectors and receivers performed exceptionally during the flight, demonstrating low photon loading and good noise performance at 280~GHz. 
The preliminary noise analysis presented here (not calibrated against the CMB) shows the best achieved combined NET for the 280~GHz instrument to be near 11.2~\mukcmbrts.
We expect that the resulting maps at 280~GHz will obtain a higher signal-to-noise on Galactic dust than any currently available in this sky region and will provide lasting value to the field.

%%%%%%%%%%%%%%%
%%%%%%%%%%%%%%%

% \disclosures 
\subsection*{Disclosures}
The authors declare that there are no financial interests, commercial affiliations, or other potential conflicts of interest that could have influenced the objectivity of this research or the writing of this paper

\subsection* {Code, Data, and Materials Availability} 
The data that support the findings of this paper are not publicly available. They can be requested from the author at elle.shaw@utexas.edu. 

\subsection* {Acknowledgments}
This work was previously published in the 2024 SPIE Millimeter, Submillimeter, and Far-Infrared Detectors and Instrumentation for Astronomy XII proceedings (doi.org/10.1117/12.3016837). The content presented here has been refined for submission to the Journal of Astronomical Telescopes, Instruments, and Systems.
\spider is supported in the U.S. by NASA under grants NNX07AL64G, NNX12AE95G, NNX17AC55G, and 80NSSC21K1986 issued through the Science Mission Directorate, and by the National Science Foundation through PLR-1043515. Logistical support for the Antarctic deployment and operations is provided by the NSF through the U.S. Antarctic Program. 
Recovery of the \spider-2 data and hardware was led by personnel from Columbia Scientific Balloon Facility, Antarctic Service Contract, \spider, Kenn Borek Air, and South Pole Station.
Support in Canada is provided by the Natural Sciences and Engineering Research Council and the Canadian
Space Agency. Support in Norway is provided by the Research Council of Norway. The Dunlap Institute is funded
through an endowment established by the David Dunlap
family and the University of Toronto. The Flatiron Institute is supported by the Simons Foundation. JEG acknowledges support from the Swedish Research Council (Reg.
no. 2019-03959) and the Swedish National Space Agency
(SNSA/Rymdstyrelsen).
We acknowledge the use of Grammarly in the final proofreading of this document.

%%%%% References %%%%%

\bibliography{report}   % bibliography data in report.bib
\bibliographystyle{spiejour}   % makes bibtex use spiejour.bst

%%%%% Biographies of authors %%%%%

\vspace{2ex}\noindent\textbf{Elle C. Shaw} is a postdoctoral fellow at the Weinberg Institute, Texas Center for Cosmology and Astroparticle Physics at the University of Texas at Austin. She received her BA in physics and mathematics from Greenville University in 2016 and her Ph.D. in physics from the University of Illinois Urbana-Champaign in 2023. Her current research interests are in CMB instrumentation, calibration, and data analysis.
She is a member of the \spider and the Simons Observatory collaborations. 

\vspace{1ex}
\noindent Biographies and photographs of the other authors are not available.

%\newpage
%\listoffigures

\end{spacing}
\end{document}